\documentclass[useAMS,usenatbib]{mn2e}
\usepackage{graphicx}

\usepackage[update,prepend]{epstopdf}

\title[Circular polarization]{Circular polarization shows the nature of pulsar magnetosphere composition}
\author[P. B. Jones]{P. B. Jones\thanks{E-mail:
p.jones1@physics.ox.ac.uk}  \\
University of Oxford, Department of Physics, Denys Wilkinson Building,\\
Keble Road, Oxford OX1 3RH, U.K.}

\begin{document}

\date{}

\pagerange{\pageref{firstpage}--\pageref{lastpage}} \pubyear{}

\maketitle

\label{firstpage}

\begin{abstract}

It has been argued in previous papers that an ion-proton plasma is formed at the polar caps of neutron stars with positive polar-cap corotational charge density.  The present paper does not offer a theory of the development of turbulence from the unstable Langmuir modes that grow in the outward accelerated plasma, but attempts to describe in qualitative terms the factors relevant to the emission of polarized radiation at frequencies below $1 - 10$ GHz.  The work of Karastergiou \& Johnston is of particular importance in this respect because it demonstrates in high-resolution measurements of the profiles of $17$ pulsars that the relative phase retardation between the O- and E-modes of the plasma is no greater than of the order of ${\rm \pi}$.  Provided the source of the radiation is at low altitudes, as favoured by recent observations, this order of retardation is possible only for a plasma of baryonic-mass particles.

\end{abstract}


\section{Introduction}

The basic features of polarization in radio pulsars were established many years ago and described in the monographs of Manchester \& Taylor (1977) and Smith (1977). Theories of its formation soon followed, with the seminal paper of Cheng \& Ruderman (1979).  More recent work has included the papers of Kazbegi, Machabeli \& Melikidze (1991), Fussell, Luo \& Melrose (2003) and Beskin \& Philippov (2012).  Of the more recent observational work, the paper of Karastergiou \& Johnston (2006) is of particular interest in that it gives the integrated polarization profiles for 17 pulsars with high resolution and good signal-to-noise ratio.  For millisecond pulsars (MSP), there are similar integrated profiles published by Yan et al (2011) and Dai et al (2015).

Two features of the integrated profiles published by Karastergiou \& Johnston are of particular interest: the general complexity and the fact that the circularly-polarized intensity though relatively small, has at most a small number of changes of hand and varies smoothly with longitude.  They are an instance of {\it res ipsa loquitur} and allow one to infer, directly and with some confidence, that the source is an ion-proton plasma accelerated outward from the polar cap.

The basis for this assertion can be outlined quite briefly. The complex profile, which is non-stochastic, and known to persist for minimum intervals of several years, must be the result of some similarly long-lived feature presumably related to the polar-cap surface. Following its emission, the observed radiation propagates through a birefringent plasma with O- and E-modes defined by linear polarizations parallel with ${\bf k}\times({\bf k}\times{\bf B})$ and ${\bf k}\times{\bf B}$, respectively, where ${\bf k}$ is the wave-vector and ${\bf B}$ the local magnetic flux density. We can make a conservative assumption that linear polarization is present at emission, then the observed longitude-dependence of the circularly-polarized intensity shows that the order of magnitude of the integrated phase difference between O- and E-modes must be no great than of the order of $\pi$.  For emission at the low altitudes that are now indicated by observation (Hassall et al 2012) this is possible only for the ion-proton plasma.  Our assumption that circular polarization is not present at emission is made because we can see no way in which the structure of a source above the polar cap could mimic terrestrial structures such as axial-mode helical antennae or coherent orthogonal dipoles.

Cyclotron absorption in the magnetosphere is negligible in an ion-proton plasm because the cross-section is inversely proportional to the mass of the particle concerned.  In the canonical secondary electron-positron plasma it is a process whose significance has been studied by a number of authors (Blandford \& Scharlemann 1976, Mikhailovskii et al 1982, Lyubarskii \& Petrova 1998, and Fussell, Luo \& Melrose 2003).  Fussell et al, in summary, remark that there is a problem in seeing how any radio-frequency emission escapes from the magnetosphere given the predicted strength of cyclotron absorption.  This process will not be considered further here but we note that the ion-proton plasma removes this dilemma.

The remaining Sections of this paper give the reasons for the assertions made above.  Section 2 gives a summary of the those assumptions about the development and decay of plasma turbulence which are needed for the emission process.  The integrated pulse profile and linear polarization are described in Section 3 and the circular polarization in section 4.  Our conclusions and a comparison with the canonical secondary electron-positron plasma models are contained in Section 5.

\section{The model system}

For a complete account of the basis for the ion-proton polar-cap model we have to refer to the sequence of papers which are cited and summarized in Jones (2015).  But this Section gives a brief summary of those points that are of specific interest here with some additional remarks.  The rapid growth of Langmuir modes is a consequence of the discrete velocities of the ion and proton beams accelerated from the polar-cap surface.  These are well-separated owing to the ion mass-to-charge ratio having values typically in the interval
 $2-3$.  The modes can be longitudinal or quasi-longitudinal (Asseo, Pelletier \& Sol 1990).  The longitudinal mode does not couple directly with the radiation field but there is no reason to suppose in a system that is not isotropic that the formation of turbulence does not provide a mechanism.
  
Whilst a study of the transition to turbulence in this system is well beyond the scope of this paper, we can compare some aspects of it with homogeneous fluid turbulence in which the distribution of energy with wavenumber is given by the Kolmogorov index (see, for example, Batchelor 1967). The immediate differences are that the accelerated particles (in the lowest Landau state) move in one dimension parallel with ${\bf B}$ and that once turbulence is established, dissipation through coupling with radiative modes must be present at all wavenumbers.  But the forces between charge density fluctuations are three-dimensional and some degree of self-similarity exists.

A further difference is that the system is expanding: the Goldreich-Julian density is $\propto \eta^{-3}$, where $\eta$ is the spherical polar coordinate radius in units of the neutron-star radius $R$. Let us consider a plasma sector of constant length $\Delta\eta$ moving outward with velocity $\sim c$ and neglect dissipation in the form of coupling with radiative modes able to propagate externally. Discounting the action of external forces, the maximum energy available for the emission of radiation is approximately the kinetic energy in its centre-of-mass frame. Charge density fluctuations relative to the local Goldreich-Julian density $\rho_{GJ}$ are $Q$ with volume $\lambda^{3}$ and bounds $Q \approx \pm\rho_{GJ}\lambda^{3}$.  The number of such elements in the sector is $N \propto \lambda^{-3}$ for $NQ$ constant and initially, they are not homogeneous as regards their composition of ions and protons.  The instantaneous electrostatic energy of the system is $\sim NQ^{2}/\lambda$.  For constant $Q$ in the expanding system, the energy $NQ^{2}/\lambda \propto \eta^{-1}$: in the limit $\eta \rightarrow \infty$ and in the absence of dissipation, electrostatic energy is converted back to kinetic energy in the centre-of-mass frame but it is disorganized rather than in the original form of two beams with well-defined velocities.
Similarly, at fixed $\eta$ but variable $Q$, $NQ^{2}/\lambda \propto \lambda^{2}$: electrostatic energy vanishes in the limit $\lambda \rightarrow 0$, as in the previous instance.  These simple considerations show that turbulence acts as a route to an energy-conserving but disordered state, with which dissipation through coupling with radiative modes is in competition.  From this, it follows that the radiation emission time-scale must satisfy $\tau^{-1}_{rad} < \eta^{-1}\partial\eta/\partial t$ for efficient generation of radio-frequency power.

There being no complete theoretical study of such a system, we are obliged to make a specific assumption about the emission process, that it is from dipole structures at causally-limited transition rates, $\tau^{-1}_{rad} \sim \nu_{c}^{4}\lambda^{3}/c^{3}$ at frequency $\nu_{c}$ in the plasma centre-of-mass frame.  Whilst both emission and absorption are present within the turbulent volume, with self-similarity in mind we assume that in the plasma centre-of-mass frame, the emission spectrum is given by,
\begin{eqnarray}
\nu_{c}N_{c}(\nu_{c}) & = & A(\nu_{c}/\nu_{c0})^{-\alpha}
\hspace{1cm}   \nu_{c} > \nu_{c0} \nonumber \\
                      & = &  0  \hspace{26mm}   \nu_{c} < \nu_{co},
\end{eqnarray}
in which $N_{c}$ is the number of photons emitted per unit volume, frequency interval, solid angle and time. The adopted value of the power-law index is $\alpha = 1.8$, consistent with the observed spectral index.  The cut-off is estimated to be typically in the interval $1 < \nu_{c0} < 10$ MHz obtained from the Langmuir mode frequency.  The radiative lifetime is then in the interval $10^{-7} <\tau_{rad} < 10^{-6}$ s and we assume that the time-scale for the development of the turbulent spectrum given by equation (1) is of the order of $10\tau_{rad}$. The important property of turbulence is that it transfers energy to higher wavenumbers. In the observer frame, the emitting region moves a distance less than of the order of $10^{-5}\gamma c \approx 3\times 10^{6}$ cm for an ion Lorentz factor $\gamma = 10$. We refer to Jones (2015) and papers cited therein giving the reasons for this order of Lorentz factor. This distance is by no means inconsistent with the compact emission volumes quoted by Hassall et al (2012) and by Kramer et al (1999) for MSP, or with the requirement for efficient generation of radio-frequency power mentioned in the previous paragraph.  Further acceleration by the Lense-Thirring effect or as a consequence of favourable flux-line curvature is possible in this interval but depends on the ion atomic number and its state of ionization, and on the whole-surface temperature of the star. 

\section{Integrated profile and linear polarization}

The integrated profile follows directly from equation (1), the flux-line geometry above the polar cap, and a Lorentz transformation to the observer frame.  The observer-frame value of the photon number $N$ is given by,
\begin{eqnarray}
N(\nu,\chi)d\nu d\Omega dt = N_{c}(\nu_{c},\zeta_{c})d\nu_{c} d\Omega_{c} dt_{c},
\end{eqnarray}
where $\chi$ and $\zeta_{c}$ are the polar angles with respect to the local ${\bf B}$ in the observer and centre-of-mass frames, respectively.
The Lorentz transformation is $\nu = \gamma\nu_{c}(1 + \beta\cos\zeta_{c})$, where $\beta$ is the source velocity.  The observer-frame energy flux obtained from equations (1) and (2) is, for an isotropic $N_{c}$,
\begin{eqnarray}
\nu N(\nu,\chi)& = &\frac{A\nu}{\gamma}\int^{\infty}_{\nu_{c0}}\frac{d\nu_{c}}{\nu_{c}}
\left(\frac{\nu_{c}}{\nu_{c0}}\right)^{-\alpha}\gamma^{2}(1 + \beta\cos\zeta_{c})^{2} \nonumber \\
&  & \delta\left(\nu_{c} - \frac{\nu}{\gamma(1 + \beta\cos\zeta_{c})}\right) \nonumber  \\
     & = & A\gamma^{2}\left(\frac{\nu}{\gamma\nu_{c0}}\right)^{-\alpha}\left(\frac{2}{1 + \gamma^{2}\tan^{2}\chi}\right)^{3 + \alpha}.
\end{eqnarray}
In the integral, $\cos\zeta_{c}$ is restricted to $-1 < \cos\zeta_{c} < 1$ and the final result assumes $\gamma \gg 1$.  In this limit, the relation between the angles is,
\begin{eqnarray}
1 + \cos\zeta_{c} = \frac{2}{1 + \gamma^{2}\tan^{2}(\chi)}.
\end{eqnarray}
The Lorentz transformation between the frequencies, $\nu = \gamma\nu_{c0}(1 + \cos\zeta_{c})$, shows that there is a minimum angle $\zeta_{c}$ and hence $\chi$ for given values of $\nu$, $\nu_{c0}$ and $\gamma$.

Radiation from a dipole element is linearly polarized, but we have to make some assumption and question whether any particular orientation might be favoured.  In a uniform one-dimensional plasma there would be none, but the plasma is not uniform: there is transverse field component ${\bf E}_{\perp}$ defined by the total acceleration potential at the given altitude and the shape of the surface separating open from closed sectors of the magnetosphere. Consequently there is a gradient $\nabla\gamma$.  The presence of some linear polarization is a basic feature of pulsar emission.  For this, but no other reason, we assume that ${\bf E}_{\perp}$ at the emission altitude defines the linear polarization of the radiation from any element in both centre-of-mass and observer frames.

Radiation on the line of sight denoted by unit vector $\hat{\bf k}$ fixed in the observer frame varies as points on the arc of traverse crossing the polar cap pass through it.  The intensity, polarization and position angle of the
radiation at any point are obtained from the three Stokes parameters $S_{0,1,2}$ for each element of polar-cap area, at the emission altitude $\eta_{e}$, by numerical integration.  An example is shown in Fig. 1 assuming a polar-cap angular radius, given by the local flux-line tangent, of
$\theta_{0}(\eta) = 0.0203\eta^{1/2}P^{-1/2}$ rad with $P = 1$ s and a cut-off $\nu_{c0} = 5$ MHz.  In order to be specific, each element is assumed to emit radiation which is completely linearly polarized.  The ion-proton plasma differs from electron-positron plasma sources in that the angle $\chi$ is significant.  The Figure shows the intensity on the arc of traverse at $50$, $150$, and $450$ MHz for Lorentz factors $\gamma = 10$ and $20$.  A minimum angle $\chi > 0$ exists if $\nu < 2\gamma\nu_{c0}$, and for both Lorentz factors allows only emission wings at $50$ MHz and much reduces the intensity at $150$ MHz.  At $450$ MHz the width if the profile is more narrow at $\gamma = 20$ than at $\gamma = 10$.  This illustrates how low-frequency wings form in a profile that is otherwise almost frequency-independent over several decades (see, for example, Hassall et al 2012).  It is also to be expected that in general, $\nu_{c0}$ will vary with position on the polar cap. (The observed width in longitude will, of course, differ from that in Fig. 1 by an unknown factor of $\csc\psi$, where $\psi$ is the magnetic orientation angle.)

\begin{figure}
\includegraphics[trim=20mm 10mm 10mm 80mm, clip, width=84mm]{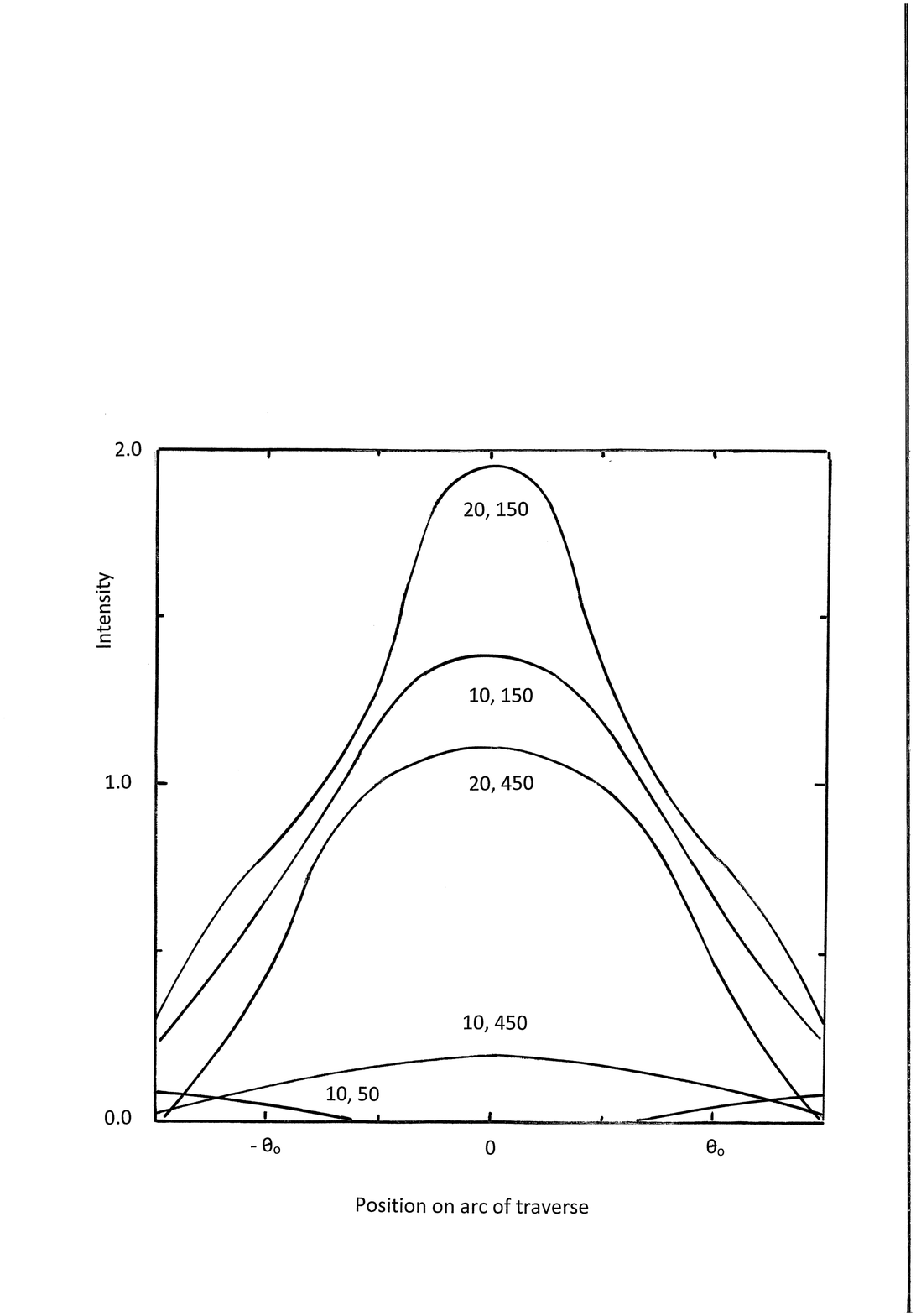}

\caption{The intensity obtained from equation (3) is given, in units of $A$, at $50$, $150$ and $450$ MHz as a function of position on the arc of traverse which is at an angular displacement of $0.2\theta_{0}$ from the magnetic axis.  For Lorentz factor $\gamma = 10$, we have $\nu < 2\gamma\nu_{c0}$ at $50$ MHz, and the cut-off at $\chi > 0$ leads to a significant intensity only in the wings.  For $\gamma = 20$, there are cut-offs at both $50$ and $150$ MHz.  The $50$ MHz intensity is too small to show on the scale of the diagram and at $150$ MHz, it is much reduced by the cut-off.  This illustrates how the cut-off (here $\nu_{c0} = 5$ MHz) in the plasma rest-frame can produce the wings that are present at low frequencies on either side of the profile.}
\end{figure}

To find the observed polarization and position angle it is necessary to make a specific choice of ${\bf E}_{\perp}$ at $\eta_{e}$.  This is shown in Fig. 2 in which the polar cap of Fig. 1 is divided into inner and outer regions.  On the semi-circle ${\bf E}_{\perp}$ is normal to the bounding surface.  The inner region has a fixed ${\bf E}_{\perp}$ in the direction that follows from the sense of flux-line curvature.  On the magnetic axis there is no curvature in a dipole field, but elsewhere the increasingly favourable curvature increases the local Goldreich-Julian charge density so producing the ${\bf E}_{\perp}$ shown.  We refer to Harding \& Muslimov (2001) for a complete analysis of this acceleration field.

\begin{figure}
\includegraphics[trim=20mm 90mm 20mm 90mm, clip, width=84mm]{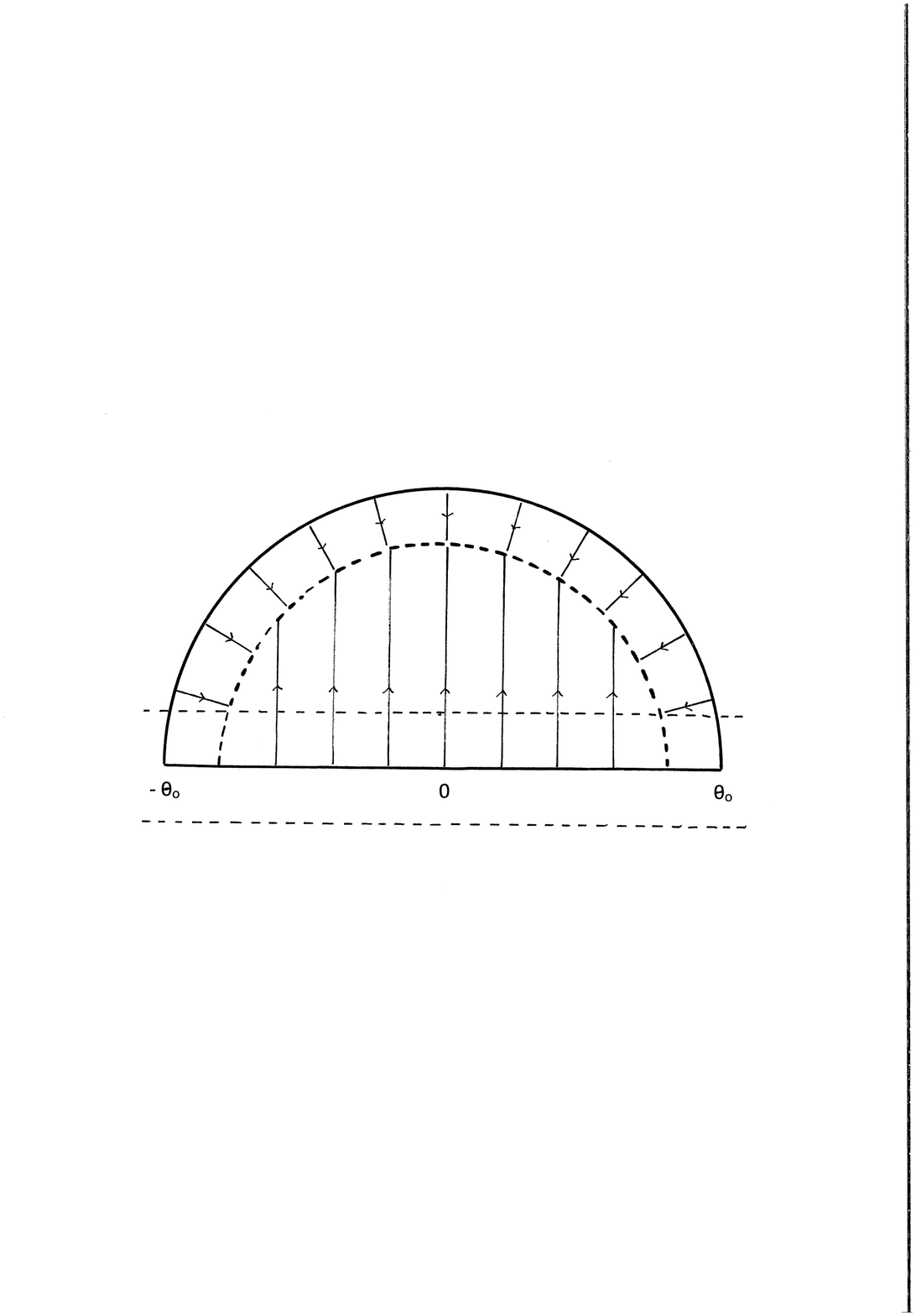}

\caption{The specific form assumed for ${\bf E}_{\perp}$ at altitude $\eta_{e}$ is shown with the polar cap and active open magnetosphere above it divided into two sectors following the arguments given in the text.  The coordinates are the spherical polar angles of the flux-line tangents at $\eta_{e}$.  This is the basis for the polarizations and position angles given in Fig. 3 for the arcs of traverse displaced by $\pm0.2\theta_{0}$ from the magnetic axis and shown in the diagram.  To be specific, radiation from individual elements of the surface of emission is assumed to be completely linearly polarized.}
\end{figure}

The simple position angle change first modelled by Radhakrishnan \& Cooke (1969) appears in only one of the pulsars studied by Karastergiou \& Johnston.  More complex behaviour is commonplace, in particular, the zeros in polarization accompanied by a position angle change of about ${\rm \pi}/2$, and a change in the sense of circular polarization which we shall discuss in the following Section. Suppose that there are two dominant incoherent sources of intensity $I_{1,2}$, polarizations $p_{1,2}$ and position angles $\phi^{p}_{1,2}$.  At a point on the arc of traverse they produce polarization,
\begin{eqnarray}
p = \frac{|I_{1}p_{1} - I_{2}p_{2}|^{2} + 2I_{1}I_{2}p_{1}p_{2}(1 + \cos(2\phi^{p}_{1} - 2\phi^{p}_{2}))}{(I_{1} + I_{2})^{2}}.
\end{eqnarray}
As is well known, this gives a zero in $p$ for $I_{1}p_{1} = I_{2}p_{2}$ and $|\phi^{p}_{1} - \phi^{p}_{2}| = {\rm \pi}/2$, but only a minimum near the point at which $I_{1}p_{1} = I_{2}p_{2}$ for position angles failing to satisfy this condition.  As points on the arc of traverse cross the line of sight $\hat{\bf k}$, the equalities of polarized intensity will often be realized. However, it is difficult to see how the position-angle orthogonality can exist except in quite small areas within the polar cap.  This is illustrated in Fig. 3 which gives the polarization and position angle (modulo ${\rm \pi}$) as functions of position on the arc of traverse for $\gamma = 100$.  Two arcs are assumed, at $\pm 0.2\theta_{0}$ with respect to the magnetic axis.  Very sharp minima are present at $-0.2\theta_{0}$ but are less prominent at $+0.2 \theta_{0}$.  The position angle jumps are quite prompt, but less so than some of those observed by Karastergiou \& Johnston, which occur within a single observation bin. This suggests that regions of relatively high Lorentz factors are necessary, restricting $\chi$ to small values.  It is possible that a different choice of polar cap, or the considerations of the last paragraph of this Section, would produce the conditions necessary for the ${\rm \pi}/2$ position angle jumps more frequently.  But it must be admitted that the observed frequency appears remarkably high.

\begin{figure}
\includegraphics[trim=15mm 50mm 5mm 55mm, clip, width=84mm]{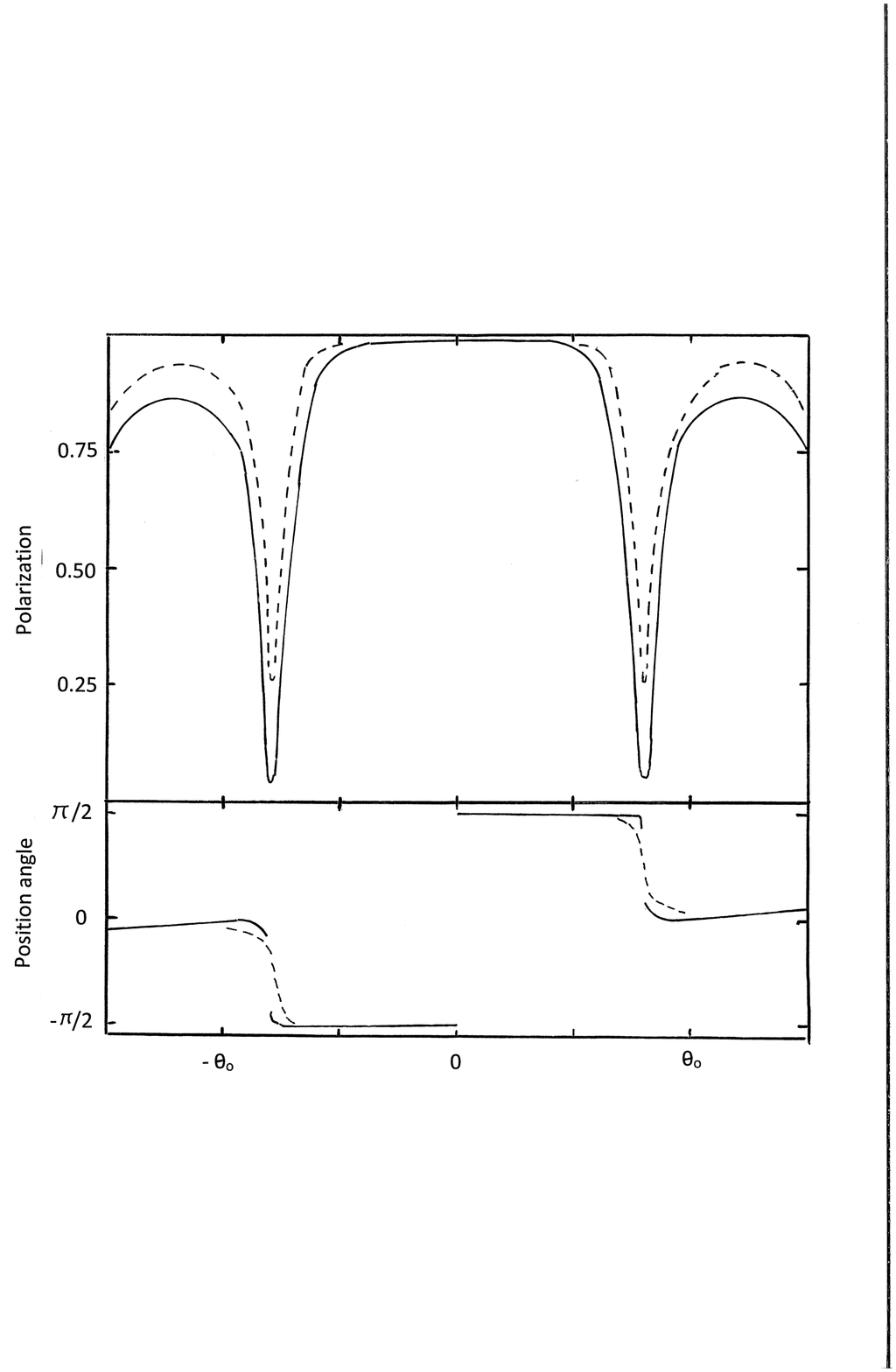}

\caption{For $1.4$ GHz and $\gamma = 100$, the polarization and position angle (modulo ${\rm \pi}$) are given as functions of position. The solid curves are for $-0.2\theta_{0}$ and the broken curves for $+0.2\theta_{0}$.  Both display sharp, deep, minima in polarization coinciding with the ${\rm \pi}/2$ jump in position angle.  There is no zero because the position angles for the two sectors are not precisely orthogonal.}
\end{figure}

The general complexity revealed by the high-resolution Karastergiou \& Johnston profiles referred to earlier is consistent with the polar-cap Goldreich-Julian charge density being positive in radio-loud pulsars, as assumed here.  The typical peak angular width is $0.04\sin\psi = 1/2\gamma$, consistent with $\gamma \sim 25$.  The detail in both intensity and polarization is not stochastic and it is extremely difficult to see how it could be the result solely of electromagnetic processes involving an electron-positron plasma subject only to space-charge-limited flow and a Dirichlet boundary condition, as would be the case in neutron stars with negative Goldreich-Julian charge density above the polar caps.

A further factor in the interpretation of the profiles is that
our choice of a semi-circular polar cap is merely an extrapolation of the asymptotic dipole field which may not be accurate even at the emission radius $\eta_{e}$.  It might be relevant that complexity in the polar-cap surface field has been studied recently in relation to the effect of Hall drift in pulsars at ages $\sim 1$ Myr (see, for example, Geppert \& Vigano 2014), which may result in an irregularly shaped boundary. There appears to be no observational evidence that the open magnetosphere at the neutron-star surface even consists of a single connected area.  In view of this, we are inclined to say that the Karastergiou \& Johnston profiles should be seen as {\it prima facie} evidence for such complexity.

\section{Circular polarization}

The previous Sections have described the production of radio frequencies at the emission altitude, which can be regarded as the surface of last absorption.  The magnetosphere at $\eta > \eta_{e}$ is certainly birefringent with separate propagation of O- and E-modes, their linear polarizations being parallel with the local values of ${\bf k}\times({\bf k \times{\bf B})}$ and ${\bf k}\times {\bf B}$ respectively. The dielectric tensor is expressed here as $\delta_{ij} + \epsilon_{ij}$, in which the $\epsilon_{ij}$ are always small compared with unity but are each a sum of
terms of identical form, one for each particle type present in the plasma.  We use the dielectric tensor given by Beskin \& Philippov (2012).  It is straightforward through the Maxwell equations to find the deviations $\Delta n$ from unity of the refractive indices for each mode.  To first order in the $\epsilon_{ij}$, we have for the O-mode,
\begin{eqnarray}
2\Delta n_{O} = \cos^{2}\theta_{k}\epsilon_{xx} + \sin^{2}\theta_{k}\epsilon_{zz} + \sin\theta_{k}\cos\theta_{k}(\epsilon_{xz} + \epsilon_{zx}),
\end{eqnarray}
and for the E-mode,
\begin{eqnarray}
2\Delta n_{E} = \epsilon_{yy},
\end{eqnarray}
in which the local direction of ${\bf B}$ is the $z$-axis and $\theta_{k}$ is the local angle between ${\bf k}$ and ${\bf B}$. A suitable approximation for the O-mode refactive index is,
\begin{eqnarray}
\Delta n_{O} = - \frac{\omega^{2}_{p}\sin^{2}\theta_{k}}{2\gamma^{3}\tilde{\omega}^{2}}.
\end{eqnarray}
Here, $\tilde{\omega} = \omega - k_{z}v^{i}_{z}$ and
$\omega^{2}_{p} = 4{\rm \pi}n_{i} {\rm e}^{2}/m_{i}$ where $v^{i}$ is the particle velocity and $i$ is the particle type, electron or proton with no distinction between ions and protons. This expression is valid for a given particle type provided $(\gamma\tilde{\omega})^{2} \ll (\omega^{i}_{B})^{2}$ and $(\gamma\tilde{\omega})^{2} \ll 4/9(\gamma\theta_{k}\omega^{i}_{B})^{2}$, where $\omega^{i}_{B}$ is the local cyclotron frequency.  The E-mode $\Delta n_{E}$ is smaller than $\Delta n_{O}$ by a factor of the order of $(\gamma\tilde{\omega})^{4}/(\omega\omega_{B})^{2}$ at angular frequency $\omega$ and can be neglected in this Section.

\begin{figure}
\includegraphics[trim=25mm 70mm 15mm 100mm, clip, width=84mm]{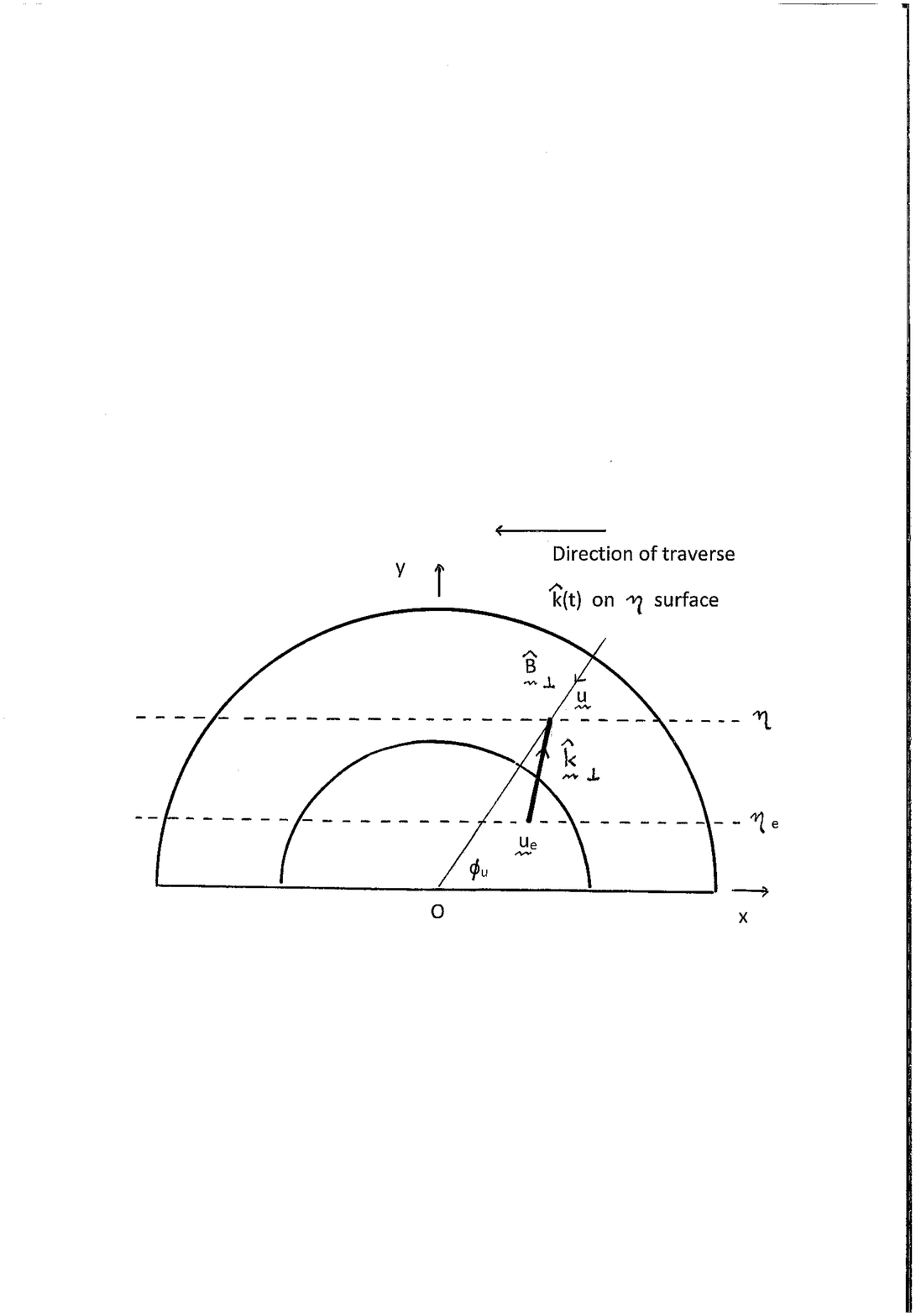}

\caption{Finding the orientation of the O- and E-mode polarizations requires the transverse vector components $\hat{\bf k}_{\perp}$ and $\hat{\bf B}_{\perp}$ at a given point ${\bf u}$ at altitude $\eta$.  The active open magnetosphere is shown at $\eta$ and at the emission altitude $\eta_{e}$ in the corotating frame of reference and here the coordinates are the spatial coordinates ${\bf u}$ and ${\bf u}_{e}$.  The line-of-sight vector $\hat{\bf k}(t)$ moves as shown from right to left.  The E-mode orientation angle, in the coordinates of the diagram, is $\phi_{E} = 0$ at the centre of a symmetrical profile: elsewhere, it has values which can be small but are dependent on the displacement of the arc of traverse from the magnetic axis.  Owing to the angle of emission $\chi > 0$, radiation from a finite area of emission, centred on the point ${\bf u}_{e}$, can have wavevector ${\bf k}$}
\end{figure}

Knowledge of $\theta_{k}$ as a function of $\eta > \eta_{e}$ is required, as well the orientation of the O- and E-modes.  To be specific, we assume dipole field geometry as shown in Fig. 4.  Here, unlike Fig. 2, the coordinates are the spatial coordinates ${\bf u}_{e}$ and ${\bf u}$ at altitudes $\eta_{e}$ and $\eta$, respectively, in the corotating frame of reference.  The time of flight between the two is $t = R(\eta - \eta_{e})/c$.  Thus the line of sight vector $\hat{\bf k}(t)$ moves as shown from right to left.  Owing to the finite value of $\chi$, radiation emitted from the vicinity of ${\bf u}_{e}$ can have direction $\hat{\bf k}$.  Equation (8) shows that $\Delta n_{O} \propto \eta^{-3}$ so that most of the birefringence arises from $\eta$ near $\eta_{e}$.  Inspection of flux-line curvature and of the angular movement of the magnetic axis in the time interval concerned show that the contributions to $\theta_{k}$ from these sources are small compared with the emission angle $\chi$ present in equation (3).  Therefore, for normal pulsars (though not for the MSP) it is a fair approximation to replace $\theta_{k}$ by $\gamma^{-1}$
for values of the local Lorentz factor no greater than $\gamma \sim 20$ and for normal pulsar periods of the order of $P = 1$ s. Under these conditions, the effects of aberration between $\eta_{e}$ and $\eta$ are also small and can be neglected.

In evaluating $\tilde{\omega}$, the difference between $\omega$ and $ck$ can be neglected so that the refractive index difference between O- and E-modes is,
\begin{eqnarray}
\Delta n = - \frac{\omega^{2}_{p}}{2\gamma\omega^{2}}.
\end{eqnarray}
This is very small, of the order of $10^{-6}$, so that refraction of the modes is negligible.  The phase retardation of the O-mode relative to the E-mode along the line of sight is,
\begin{eqnarray}
\phi_{ret} = \frac{R\omega}{c}\int^{\infty}_{\eta_{e}}\Delta n_{O}(\eta)d\eta   
 =    \frac{R\omega^{2}_{p}(1)}{4\gamma\omega c \eta_{e}^{2}},
\end{eqnarray}
in terms of $\omega_{p}$ for the particle type in question, evaluated at the neutron-star surface.

The instantaneous components of $\hat{\bf B}$ and $\hat{\bf k}$ perpendicular to the magnetic axis at the point where radiation is at altitude $\eta$ are shown in Fig. 4 and from them, in the small angle approximation, we can follow the evolution of the E-mode polarization along the arc of traverse.  In the small-angle approximation ($\hat{k}_{\perp}, \hat{B}_{\perp} \ll 1$) it is given by,
\begin{eqnarray}
\tan\phi_{E} = \frac{\hat{k}_{x} - a\cos\phi_{u}}{a\sin\phi_{u} - \hat{k}_{y}}, 
\end{eqnarray}
with $a = 3u/2\eta R$.  Thus $\phi_{E} = 0$ at the centre of a symmetrical profile.  Otherwise, it is generally oriented, may be small, and is dependent on $\hat{\bf k}$.

The effect of plasma on the radiation can be summarized as follows.

(i)	  The radio frequency is formed at altitude $\eta_{e}$ with linear polarization, as described in Section 3, which can be resolved into a a linear combination of the O- and E-modes, their relative amplitudes depending on the angle $\phi_{lin}$ between the formation electric vector and that of the O-mode found via equation (11).  This angle tends to be small in dipole field geometry, in which case the O-mode is the dominant component.

(ii)  Radiation propagates at $\eta > \eta_{e}$ with a small adiabatic rate of change in its polarization orientation given by equation (11), referred to as adiabatic walking (see Cheng \& Ruderman 1979).  The refractive index given by equation (9) is so small for an ion-proton plasma that there is no question of coherence loss through spatial separation of the modes.

(iii)  The rate of increase in $\phi_{ret}$ becomes negligible at some $\eta = \eta_{lp}$, which is the polarization limiting radius.

Cyclotron absorption occurs at some $\eta > \eta_{lp}$ but the rate is negligible as is its contribution  to $\Delta n$. In the case that $\phi_{lin}$ is small, the polarization at $\eta_{lp}$ consists of a linear component little changed by the birefringence accompanied by a relatively small circular component.  In terms of the Stokes parameters, $S^{2}_{3} \ll S^{2}_{1} + S^{2}_{2}$.	 The sequence (i) - (iii)  is completely analogous with the passage of light through a quarter-wave plate in optics, except that in that case, $\phi_{lin} = \pi/4$ is chosen so that the O- and E-modes have equal amplitude.

The existence of circular polarization depends on how the linear polarization at $\eta_{e}$ is represented as a linear superposition of the two plasma modes.  Population of only one mode produces a zero in the circular polarization.  Therefore, in the traverse of a polar cap having some degree of symmetry, the source linear polarization would be expected to populate only the O-mode at the intensity profile peak, so producing a zero with a change of sense in the observed circular polarization.  This is a well-known phenomenon and is clearly visible in 6 of the 17 pulsars studied by Karastergiou \& Johnston.

In the previous Section, the zeros in linear polarization coinciding with ${\rm \pi}/2$ jumps in position angle were described.  A zero in linear polarization must be accompanied by a zero in circular polarization for any value of $\phi_{ret}$ if the circular polarization is simply a consequence of birefringence external to the source.  The change of sense follows from the ${\rm \pi}/2$ change in orientation of the linear polarization.

The preceding remarks serve to show that our view of circular polarization is by no means inconsistent with phenomena which have been observed now for almost half a century. The primary purpose of this paper is to place equation (10) in context and to evaluate it, choosing the most usual observing frequency of $1.4$ GHz and a Goldreich-Julian number density. For the proton mass, and for $R = 1.2\times 10^{6}$ cm, we find,
\begin{eqnarray}
\phi_{ret} = 136\frac{B_{12}}{P}\frac{1}{\gamma\eta^{2}_{e}},
\end{eqnarray}
where $B_{12}$ is the surface magnetic flux density in units of $10^{12}$ G.
We have no means of knowing precise values of $\gamma$ and $\eta_{e}$ for any particular pulsar, but equation (12) indicates that, for the Karastergiou \& Johnston pulsars excepting J0835-4510, we can expect the order of magnitude $\phi_{ret} \sim {\rm \pi}$ for $\gamma = 20$ and $\eta_{e} =4$.  Evaluation of equation (10) for electrons, again for a Goldreich-Julian density, would simply yield equation (12) multiplied by $1836$, the proton-electron mass ratio.

The question is, how would such large values of $\phi_{ret}$ appear in circular polarization profiles?  If birefringence at $\eta > \eta_{e}$ is the only significant mechanism, the ratio of circular polarization to linear polarization intensities at any longitude is $\sin^{2}\phi_{ret}\sin^{2}\phi_{lin}$ in which $\phi_{lin}$ is the angle between the source linear polarization at $\eta_{e}$ and either of the local O- and E-mode axes given by equation (11).  The fact that $\phi_{lin}$ may be small explains why the observed ratio varies between such wide limits.  The important point is that the parameters  $\gamma$ and $\eta_{e}$ must vary with observation longitude.  Thus $\sin\phi_{ret}$ would be certain to vary rapidly with longitude if its absolute magnitude were a large multiple of $2{\rm \pi}$. But a change $\phi_{ret} \rightarrow \phi_{ret} + {\rm \pi}/2$ changes the circular polarization intensity from a maximum to zero, or vice-versa, and a change of ${\rm \pi}$ changes its sense. The Karastergiou \& Johnston profiles constructed from $1024$ bins per pulse period provide no evidence for such behaviour.  Indeed, the leisurely movement of the degree of polarization in general, and particularly through zero with change of sense that is observed in profile centres, is consistent with $\phi_{ret}$ having the modest values predicted for the ion-proton plasma.

It might be argued that, for the canonical secondary electron-positron plasma, the predicted values of $\phi_{ret}$ could be much reduced by increasing $\eta_{e}$, approximately by the cube root of the proton-electron mass ratio multiplied by the number of pairs per primary beam particle.  But high altitudes are not favoured by recent LOFAR observations of normal pulsars (see Hassall et al 2012) and would be in the vicinity of MSP light cylinders.

The possibility of a small secondary electron-positron component being present in the ion-proton plasma was considered at length in Jones (2015) owing to its effect on the rate of growth of Langmuir modes.  Positrons accelerated to energies of $10^{4-5}$ Mev are not significant in this respect: their Lorentz factors are too great for them to participate other than negligibly in collective plasma modes.  Here, this parameter in equation (8) would render them insignificant as would the Lorentz transformation to the rest frames of inward-moving photo-electrons from the accelerated ions.

This Section has so far not considered circular polarization in MSP spectra.
The principal reason is that evaluation of $\Delta n_{O}$ would be more complex than the simple result given by equation (8).  The rotational angular velocities result in factors such as aberration and flux-line curvature being significant in the estimation of $\theta_{k}$.  The cyclotron frequencies would be so low that the inequalities necessary for equation (8) could not be satisfied.  Yan et al (2011) observe at $1.4$ GHz the same level of profile complexity as Karastergiou \& Johnston.  Bearing in mind that observations (Kramer et al 1999) favour a very compact emission zone in MSP, there appears to be no reason to suppose that, at least qualitatively, the normal pulsar results of this Section are not valid for the MSP.

\section{Conclusions}

Sections 1 - 3 have attempted to set out those features of the ion-proton plasma that lead to coherent linearly-polarized radio-frequency emission.  Langmuir-mode growth rates are such that non-linearity and a transition to turbulence can occur at altitudes in the interval $2 < \eta < 10$ above the polar cap (see Jones 2012, 2015).  Following this, it has been argued that turbulent development and  coherent emission occur at low altitudes and in intervals at least as compact as those favoured by observation.

The reasons why the radiation at the formation altitude, which for simplicity we denote by the unique value $\eta_{e}$, should be partially linearly-polarized are less compelling.  It must be admitted that this is an assumption, but with the merit that it is not inconsistent with observation.

The plasma at $\eta > \eta_{e}$ on the line of sight is outward-moving and, whatever its composition, is birefringent in the sense of having different refractive indices for its O- and E-modes and its effect on radiation propagation in the interval immediately beyond $\eta_{e}$ can be estimated easily. It is argued in Section 4 that the circular-polarization profiles published by Karastergiou \& Johnston are consistent only with the absolute value of the O-mode retardation phase being not more than of the order of ${\rm \pi}$.  Given the low formation altitude $\eta_{e}$, this is consistent with an ion-proton plasma of Goldreich-Julian density having little low-energy positron contamination. Owing to the proton-electron mass ratio and to its multiplicity per unit primary beam particle, the canonical secondary electron-positron plasma would produce an absolute retardation phase many orders of magnitude too large.   This conclusion is simple and appears robust as it depends essentially only on the emission altitude $\eta_{e}$ and the interpretation of the Karastergiou \& Johnston profiles.  We emphasize that it  is independent of  the detail discussed in Sections 2 and 3.  It is also independent of the presence of intrinsic circular polarization at source, $\eta < \eta_{e}$.  If low-energy electrons are present at Goldreich-Julian densities or above, their contribution to the phase retardation will be large and cannot be cancelled out.

We have stated in Section 1 that cyclotron absorption is negligible in an ion-proton plasma, whereas in a high-multiplicity electron-positron plasma, high absorption would be expected in certain circumstances.  We are unaware that there is any unambiguous published evidence for its presence in radio-loud pulsar profiles or spectra.

Our principal conclusion here is that radio-loud pulsars with the common large negative spectral indices are neutron stars with positive polar-cap corotational charge densities.  Basic nuclear physics reactions ensure that these neutron stars have an unstable outward flux of ions and protons from the polar cap.  This view has been published previously (Jones 2013) as a consequence of the LOFAR emission height measurements of Hassall et al.  The observability of isolated neutron stars with negative polar-cap charge density is then questionable.  Certainly they should have incoherent emission at early times after formation but the standard coherent radio emission is not expected.  Some other form of coherent emission may exist, possibly of weak intensity or with a small negative spectral index.  Future observations with the Square Kilometre Array or at high GHz frequencies should be interesting.

\section*{Acknowledgments}

I thank Aris Karastergiou and Simon Johnston for a very stimulating discussion about the significance of their work on polarization.

\bsp

\label{lastpage}

\end{document}